# Photonic crystal nanocavity laser with a single quantum dot gain

Masahiro Nomura, Naoto Kumagai, Satoshi Iwamoto, Yasutomo Ota, and Yasuhiko Arakawa

*Institute for Nano Quantum Information Electronics, Institute of Industrial Science, The University of Tokyo, Tokyo 153-8505, Japan*

nomura[at]iis.u-tokyo.ac.jp

We demonstrate a photonic crystal nanocavity laser essentially driven by a self-assembled InAs/GaAs single quantum dot gain. The investigated nanocavities contain only 0.4 quantum dots on an average; an ultra-low density quantum dot sample ($1.5 \times 10^8$ cm$^{-2}$) is used so that a single quantum dot can be isolated from the surrounding quantum dots. Laser oscillation begins at a pump power of 42 nW under resonant condition, while the far-detuning conditions require ~145 nW for lasing. This spectral detuning dependence of laser threshold indicates substantial contribution of the single quantum dot to the total gain. Moreover, photon correlation measurements show a distinct transition from anti-bunching to Poissonian via bunching with the increase of the excitation power, which is also an evidence of laser oscillation with the single quantum dot.

Semiconductor nanocavity systems with a self-assembled single quantum dot (QD) [1] have been investigated because of their unique physics based on cavity quantum electrodynamics and their potentials in future applications such as quantum information processing. In semiconductor microcavity systems [2], vacuum Rabi splitting in the strong-coupling regime [3-6] and highly efficient lasing in the weak-coupling regime [7-15] have been observed. The laser oscillation in semiconductor microcavity systems that contain only a single QD is currently of considerable interest. Thus far, microcavity lasers with single QD gain have been fabricated using microdisk [8] and micropillar [9] structures. These microcavity systems contained tens or hundreds of QDs per cavity. Therefore, interference arises not only from the target QD but also from other QDs, hindering access to the delicate physics of a single QD-cavity system. This deviation in behavior from an isolated quantum system can be minimized by employing a small cavity in a wafer with an extremely low areal density of QDs. Due to the small mode volume and high cavity quality factor ($Q$), the use of a photonic crystal (PhC) nanocavity [16, 17] is one of the most promising approaches for investigating physics of single QD-cavity systems. We have fabricated nanocavity systems with highly isolated single QDs (~ 0.4 QD/cavity), using a small PhC nanocavity and an ultra-low density QD sample.

In this study, we demonstrate a PhC nanocavity laser essentially driven by a single QD. Distinct single QD features are observed during gain tuning measurements and photon correlation measurements. The cavity coupled SQD provides dominant gain (>70%) of the system. The photon statistics change from anti-bunching to Poissonian via bunching, which indicates a phase transition, as the excitation power is increased.

Self-assembled QDs were grown by molecular beam epitaxy on an undoped (100)-oriented GaAs substrate. First, a 300-nm-thick GaAs buffer layer was deposited on the substrate. Then, a 700-nm-thick $Al_{0.6}Ga_{0.4}As$ sacrificial layer was grown. Finally, a 160-nm-thick GaAs slab layer along with a single self-assembled InAs QD layer was grown at the center of the substrate. The photoluminescence (PL) peak of the QD ensemble was observed at 930 nm at 6 K. The nominal areal QD density was ~1.5 x $10^8$ cm$^{-2}$. PhC microstructures were fabricated by electron beam lithography, inductively coupled plasma reactive ion etching, and a wet etching process using a hydrofluoric acid solution, which formed 160-nm-thick air-bridge structures by removing the sacrificial layer. We fabricated a sample with a period of the lattice $a$ = 235 nm and radius of the air hole $r$ ~0.28 $a$. Triangular lattice air holes were patterned using an electron beam lithography system and an inductively coupled plasma reactive ion etching process. Finally, the AlGaAs sacrificial layer was removed to form the air bridge structures. This series of processes were used to fabricate a semiconductor based air-bridged PhC slab with an air-hole array, which produces an in-plane photonic bandgap. We adopted a point defect structure, called L3 defect, which consists of three missing air holes along the Γ-K direction of the triangular PhC lattice. In addition, the first and third nearest air holes at both edges of the cavity were shifted to outside the cavity to obtain higher cavity quality factor $Q$ [18]. The displacement of the shifted air holes was 0.16$a$. This structure confines photons within an extremely small mode volume of $V_m \sim 0.7(\lambda/n)^3$, as shown in the lower right inset of Fig.1; the system was simulated using a finite-difference time-domain method. Here, $\lambda$ denotes the wavelength of the cavity mode in vacuum and $n = 2.9$ is the effective refractive index. The mode volume of the laser was then calculated to be ~0.02 μm$^3$. Thus, our PhC nanocavity has the essential advantage of spatially filtering the number of QDs within the cavity.

The measurements were performed at cryogenic temperature using a micro-photoluminescence (μ-PL) setup. A CW Ti:Sapphire laser operated at 800 nm was used an excitation source. An excitation beam was focused on the surface of the sample using a microscope objective lens (50x, numerical aperture = 0.42) in the normal direction, and positioned on the PhC using piezoelectric nanopositioners. The theoretical diameter of an excitation spot formed on the surface of the sample was calculated to be ~2.3 μm, which was smaller than that of the PhC pattern. The PL was collected by the same microscope objective lens as that used for focusing the excitation beam.

We adopted a PhC nanocavity and an ultra-low density QD sample to reduce an average number of QDs in the cavity. The areal density of self-assembled InAs QDs in our semiconductor wafer ranges from 1 - 2 per μm$^2$. Therefore, the average number of QDs in the cavity is only 0.4, which is more than two orders of magnitude smaller than that used in previous investigations [8, 9]. Furthermore, the $\delta$-function-like density of states of a QD further reduces the average number of QDs by spectral filtering, minimizing the degree of interference. The measured PL spectrum at 6 K (Fig. 2(a)) consisted of a single exciton (red line) and cavity mode (blue line, estimated $Q$ ~25,000).

The exciton-mode coupling in our system was finely controlled using a temperature-tuning technique, in which an exciton line is scanned through the cavity resonance as shown in Fig. 2(b). This technique tunes the relative spectral positions of a target QD and the fundamental cavity mode, based on the different temperature dependence of the bandgap and of the refractive index. Photoluminescence (PL) spectra were recorded at an irradiated pump power (defined as the power at the sample surface) of ~60 nW as a function of the detuning between the cavity mode and target exciton, $\Delta\lambda = \lambda_x - \lambda_m$, where $\lambda_x$ and $\lambda_m$ denote the wavelengths of the target exciton and cavity mode, respectively. The spectra were measured in the vicinity of zero detuning. The sample temperature was controlled between 27 K and 45 K in order to vary the value of $\Delta\lambda$ between −0.4 nm and 0.4 nm. No significant optical degradation of the exciton was observed on increasing the temperature in this range. At this pump power, lasing occurs only at zero detuning due to the sharp excitonic gain spectrum, indicating that the single exciton plays an essential role in the laser oscillation.

The contribution of the coupled, single QD gain to the laser oscillation was quantitatively investigated by measuring the laser threshold at various detunings. Figure 3(a) shows PL spectra measured under resonant (red) and far-detuning (blue) conditions. The coupling of a single QD drastically enhances the intensity of the cavity mode. Figure 3(b) shows light-in versus light-out (L-L) plot collected at $\Delta\lambda = 0$ nm. At zero detuning, the threshold was estimated to be ~42 nW, while sufficiently detuned cases required ~145 nW on average (Fig. 3(c)). Thus, the coupling of the single exciton significantly increases the material gain of the system, and results in a significant reduction of the threshold pump power compared with the far-detuning condition. We can estimate that the dominant gain (~71%) is supplied by the coupling single QD. Lasing was observed even under the far-detuning condition. We investigated the gain source of the cavity mode by carrying out cross-correlation measurement under a far-detuning condition of $\Delta\lambda \sim -3.7$ nm. The observed cavity-exciton anti-correlation indicates the occurrence of non-resonant coupling between the single QD and the cavity mode and subsequent non-resonant lasing. This unidentified channelling mechanism is caused by several factors including phonon interaction processes [7, 19, 20]. This channelling mechanism enables the single QD to provide the gain in the far-detuning conditions; a net single QD gain may be larger than 71%.

When the Pauli blocking in the system is negligible, the L-L plot of a single QD laser and that of a multi-QD laser are not significantly different. However, a unique characteristic of such a laser is observed in photon statistics. To investigate the quantum-statistical characteristics of the photon stream from the laser, we measured the photon correlation function $g^2(\tau) = \langle I(t)I(t+\tau)\rangle / \langle I(t)I(t)\rangle$ under the coupled exciton condition ($\Delta\lambda = 0$) using a Hanbury Brown-Twiss setup (Fig. 4(a)) [21]. Here, $\langle I(t) \rangle$ is the expectation value of the intensity of the laser at time $t$, and $\tau$ denotes a delay time. Photon coincidences were recorded electronically in the form of a histogram of start-stop events using two avalanche photodiode single-photon counters. An example of a measurement of $g^2(\tau)$ below the threshold (0.34$P_{th}$, where $P_{th}$ is the threshold pump power of 42 nW) is given in Fig. 4(b). This measurement demonstrates that the light emitted from the single-exciton coupled laser is manifestly non-classical, exhibiting photon anti-bunching $g^2(0) < g^2(\tau)$ and sub-Poissonian photon statistics $g^2(0) = 0.42 < 1$ [22]. This anti-bunching behavior, the suppression of multi-photon emission, is a well-known feature of single photon sources that use a single QD [23]. We estimated that ~76% of the photons in the cavity mode were present as a consequence of the single QD on resonance. This value is in good agreement with the single QD contribution of 71%, which is estimated in Fig. 3(c). Therefore, our observations indicate that the laser exhibits distinct characteristics of a single QD in the low power excitation regime.

Also of significance is the observation of enhanced and stabilized quantum noise near and above the threshold, respectively. Figure 4(c) shows a plot of $g^2(\tau)$ recorded near the threshold pump power ($1.35P_{th}$), which exhibits a strongly enhanced multi-photon coincidence probability in the vicinity of delay time $\tau = 0$. This type of photon-bunching behavior close to the laser threshold is a well-known characteristic of conventional lasers [24], cavity-quantum electrodynamics lasers [25], and even thresholdless lasers [26]. In contrast to this non-classical and enhanced-amplitude noise feature, stabilization of the intensity noise was observed in the high power excitation regime ($9.3P_{th}$, Fig. 4(d)). The flat $g^2(\tau)$ trace indicates that the photon statistics is Poissonian (that is, coherent), implying that laser oscillation occurs. The demonstrated laser action verifies the presence of distinct positive correlation during the phase transition from single-photon emission regime to laser oscillation regime.

In order to investigate the laser further, the photon correlation function was recorded over a wide range of pump powers and was analyzed by fitting to the function $g^2(\tau) = 1 - (1 - g^2(0))\exp(-|\tau|/\tau_0)$, using two fitting parameters, $g^2(0)$ and $\tau_0$; the latter is the decay time of the dominant or combined dynamics in the system. The dominant physics varies as the pump power is increased through the laser threshold. The logarithmic L-L plot in Fig. 4a reveals that the pump power can be classified into three regimes: a spontaneous emission regime (light blue), a phase transition regime (orange), and a lasing regime (pink). The experimental data (blue spheres) take the form of a gentle s-shaped curve. A smooth transition such as this from the spontaneous to stimulated emission region is typically observed for lasers in which the spontaneous emission efficiently couples to the lasing mode. The experimental curve was fitted (light blue) using conventional coupled rate equations for the carrier density and the photon density [14, 27]. The spontaneous emission coupling factor was estimated to be approximately 0.4. The physics around the laser threshold, in the region $0.5P_{th} < P < 2P_{th}$, is significant. The observed photon bunching, with $g^2(0) > 1$, is a manifestation of the enhancement in the noise amplitude at the threshold, where the spontaneous and stimulated emission processes coexist and are comparable in importance. This photon bunching may be caused by the coupling of photons emitted by other unidentified background oscillators. The gradual increase of $g^2(0)$ with pump power below the threshold can be explained by a change in the dominant photon statistical feature from single emitter-like to laser-like at the lasing threshold. Well above the threshold, in the region $P > 2P_{th}$, $g^2(0)$ gradually decreases with pump power and eventually approaches unity (dashed green line). This behavior indicates that stimulated emission dominates the photon emission process here and that the system has reached the lasing regime.

In summary, a photonic crystal nanocavity laser with single quantum dot gain was demonstrated. A small cavity and an ultra-low density quantum dot sample resulted in high isolation of a target single QD (~0.4 QD/cavity). A QD exciton tuning measurements indicated that substantial contribution of the single QD to the total gain. In photon correlation measurements, photon bunching was observed during the phase transition from single photon emission to coherent light generation regime.

We thank S. Ishida, M. Shirane, S. Ohkouchi, Y. Igarashi, S. Nakagawa, and K. Watanabe for their technical support. We thank T. Nakaoka, A. Tandaechanurat, and S. Kako for fruitful discussions. This research was supported by the Special Coordination Funds for Promoting Science and Technology and by Kakenhi 20760030, the Ministry of Education, Culture, Sports, Science and Technology, Japan.

# Figures

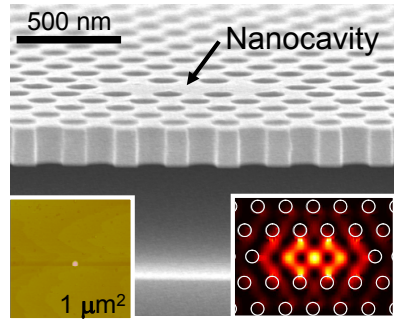

Fig. 1. Scanning electron micrograph of the PhC nanocavity laser. An atomic force microscope image of an equivalent sample without capping demonstrates that no interference from other quantum dots occurs (lower left inset). The lower right inset depicts the electric field intensity of the cavity mode, showing that photons are strongly confined.

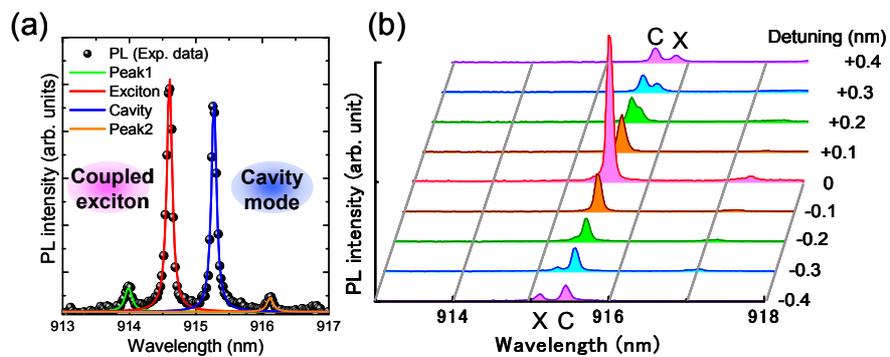

Fig. 2. (a) PL spectrum of the coupled exciton (914.6 nm) and the cavity mode (915.25 nm) at sufficiently high detuning (6K). (b) PL spectra recorded at various detunings for a pump power of 60 nW; x and c denote the exciton and the cavity, respectively.

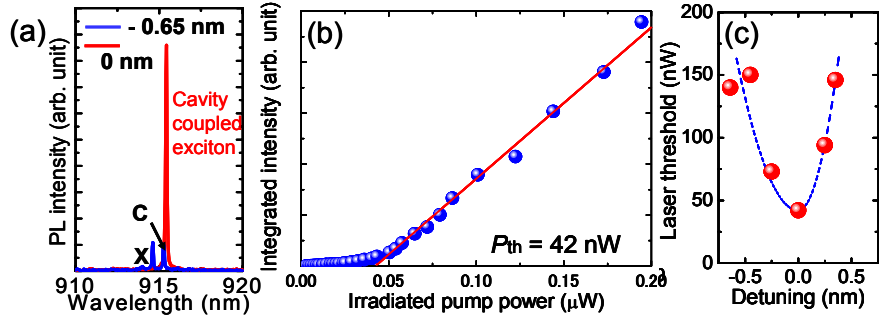

Fig. 3. (a) PL spectra measured under far-detuning (-0.65 nm, blue) and resonant (red) conditions at the excitation power of ~15 nW. (b) L-L plots of the cavity mode under coupled condition. (c) Spectral detuning dependence of laser threshold. Lasing begins at a pump power of 42 nW under resonant condition, while the far-detuning conditions require ~145 nW for lasing. The detuning of the single QD to the cavity mode was carried out by changing the temperature.

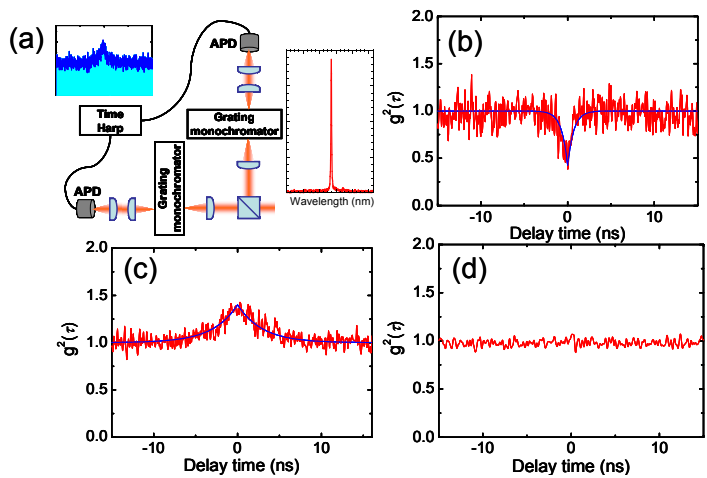

Fig. 4. Photon correlation measurements for a laser with a single QD under coupling condition. (a) Schematic picture of the optical system used in the measurements. (b)-(d), Photon correlation function $g^2(\tau)$ recorded at below (0.34$P_{th}$), near (1.35 $P_{th}$), and above (9.3 $P_{th}$) the laser threshold ($P_{th}$ = 42 nW) under the condition of zero detuning. The photon statistics changes from anti-bunching (b) to bunching (c) to Poissonian (d) as the pump power is increased. The blue lines in (b) and (c) are fitted curves.

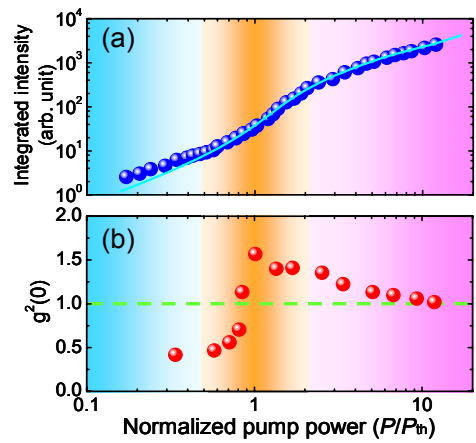

Fig. 5. Photon statistics of a single quantum dot coupled laser. (a), L-L plot on a logarithmic scale with the fitted curve shown in light blue. (b), Photon correlation function $g^2(0)$ at various pump powers. The horizontal axes of the two panels represent the pump power normalized by Pth = 42 nW. The dashed green line $g^2(0) = 1$ indicates the photon statistics of coherent light. The change in $g^2(0)$ clearly shows a transition of the light source from a single photon source to a laser with an enhancement of the intensity noise at the threshold.